\documentclass[%
preprint,
superscriptaddress,
showpacs,
 amsmath,
 amssymb,
aps,
pre,
floatfix,
]{revtex4-1}
\usepackage{graphicx}
\usepackage[colorlinks,allcolors=blue]{hyperref}
\usepackage{amsmath,amssymb,amsthm}

\usepackage{array, booktabs, makecell}

\begin{document}

\title{Quenched and Annealed Disorder Mechanisms in Comb-Models with Fractional Operators}

\author{A. A. Tateishi}
\affiliation{Departamento de F\'{\i}sica, Universidade Tecnologica Federal de Pato Branco, Pato Branco}

\author{H. V. Ribeiro}
\affiliation{Departamento de F\'{\i}sica, Universidade Estadual de Maring\'a - Maring\'a, PR 87020-900, Brazil}

\author{T. Sandev}
\affiliation{Research Center for Computer Science and Information Technologies, Macedonian Academy of Sciences and Arts, Bul. Krste Misirkov 2, 1000 Skopje, Macedonia}
\affiliation{Institute of Physics \& Astronomy, University of Potsdam, D-14776 Potsdam-Golm, Germany}
\affiliation{Institute of Physics, Faculty of Natural Sciences and Mathematics, Ss Cyril and Methodius University, Arhimedova 3, 1000 Skopje, Macedonia}

\author{I. Petreska}
\affiliation{Institute of Physics, Faculty of Natural Sciences and Mathematics, Ss Cyril and Methodius University, Arhimedova 3, 1000 Skopje, Macedonia}

\author{E. K. Lenzi}\email{eklenzi@uepg.br}
\affiliation{Departamento de F\'{\i}sica, Universidade Estadual de Ponta Grossa, Av. Carlos Cavalcanti 4748, 84030-900~Ponta~Grossa, PR, Brazil}

\linespread{1.3}
\begin{abstract}
Recent experimental findings on anomalous diffusion have demanded novel models that combine annealed (temporal) and quenched (spatial or static) disorder mechanisms. The comb-model is a simplified description of diffusion on percolation clusters, where the comb-like structure mimics quenched disorder mechanisms and yields a subdiffusive regime. Here we extend the comb-model to simultaneously account for quenched and annealed disorder mechanisms. To do so, we replace usual derivatives in the comb diffusion equation by different fractional time-derivative operators and the conventional comb-like structure by a generalized fractal structure. Our hybrid comb-models thus represent a diffusion where different comb-like structures describe different quenched disorder mechanisms, and the fractional operators account for various annealed disorders mechanisms. We find exact solutions for the diffusion propagator and mean square displacement in terms of different memory kernels used for defining the fractional operators. Among other findings, we show that these models describe crossovers from subdiffusion to Brownian or confined diffusions, situations emerging in empirical results. These results reveal the critical role of interactions between geometrical restrictions and memory effects on modeling anomalous diffusion.
\end{abstract}
\maketitle

\section{Introduction}

The development of the diffusion concept has always relied on the mutually-beneficial relationship between theory and experiments. Since Perrin's experiments proving Einstein's diffusion theory~\cite{Perrin1910}, Brownian (usual) diffusion is well-known to display a Gaussian distribution and a linear time dependence of the mean square displacement (MSD). However, as deviations from these usual behaviors started to appear in experimental studies of disordered media and biological systems, the need to understand underlying microscopic mechanisms of these unusual dynamics has given rise to breakthrough theories in statistical physics. 

The anomalous diffusion era started with the concept of waiting-time distribution in random walks, proposed independently by L\'evy~\cite{Levy1954}, Smith~\cite{Smith1955}, and Montroll and Weiss~\cite{Montroll1965}; but the continuous-time random walk (CTRW) emerged as the foundation of anomalous transport only after the works of Scher and Lax~\cite{Scher1973a,Scher1973b} on unusual results for charge transport in amorphous semiconductors. These works use CTRW to describe heterogeneities of a medium in an annealed way, where the waiting-time distribution represents the environment randomness. The ``second youth'' of the CTRW~\cite{Masoliver2017} is usually marked by its relationship with fractional diffusion equations~\cite{Klafter1987,Hilfer1995,Compte1996,Barkai2002}, a formalism that becomes known as an efficient phenomenological description of anomalous diffusion in complex systems~\cite{Metzler2000,West2016,Masoliver2017}.

Percolation theory~\cite{Broadbent1957} represents another significant breakthrough for the description of anomalous diffusion. As emphasized by Havlin and Ben-Avraham~\cite{Havlin2002}, the percolation model is a simple and purely geometrical approach to describe disordered media. While random walk processes are generalized in the CTRW model, in percolation processes, usual random walks take place in a disordered environment. In contrast to the CTRW framework, percolation theory thus describes a quenched disorder, where the randomness associated with geometrical constraints are constant in time. Under this context, de Gennes~\cite{deGennes1976} coined the term ``the ant in a labyrinth'' to describe random walks in percolation lattices and established a paradigm of anomalous diffusion caused by geometrical structures~\cite{Bouchaud1990,Bok2009}. This paradigm becomes well established mainly due to fractal geometry~\cite{Mandelbrot1989}, and it is essential in the study of porous media~\cite{Havlin2002}.

Diffusion also becomes an essential noninvasive tool to probe and characterize systems ranging from materials to living organisms~\cite{Sen2004,Waigh2005,Kirstein2007,Wirtz2009,Novikov2014,Papaioannou2017,Assaf2019,Song2019}. These recent empirical results revealed a myriad of complex patterns that are usually not well described by analytical tools developed for amorphous solids and porous media. As argued by Metzler~\cite{MetzlerBiophysJ2017}, these novel experimental findings require researchers to come up with novel models. In this context, hybrid or mixed models of anomalous diffusion emerged as a significant modeling possibility~\cite{Meroz2010,Weigel2011,Jeon2011,Tabei2013,Miyaguchi2015,Golan2017,Furnival2017}. Examples include CTRW on fractals~\cite{Meroz2010,Weigel2011,Golan2017}, CTRW combined with fractal Brownian motion~\cite{Jeon2011,Tabei2013}, quenched-trap model and fractal lattices~\cite{Miyaguchi2015}, and CTRW combined with percolation theory~\cite{Furnival2017}. In general, these models share the idea of combining annealed (temporal) and quenched (spatial or static) disorder mechanisms. 

\begin{figure}[!t]
\centering
\includegraphics[width=0.9\textwidth]{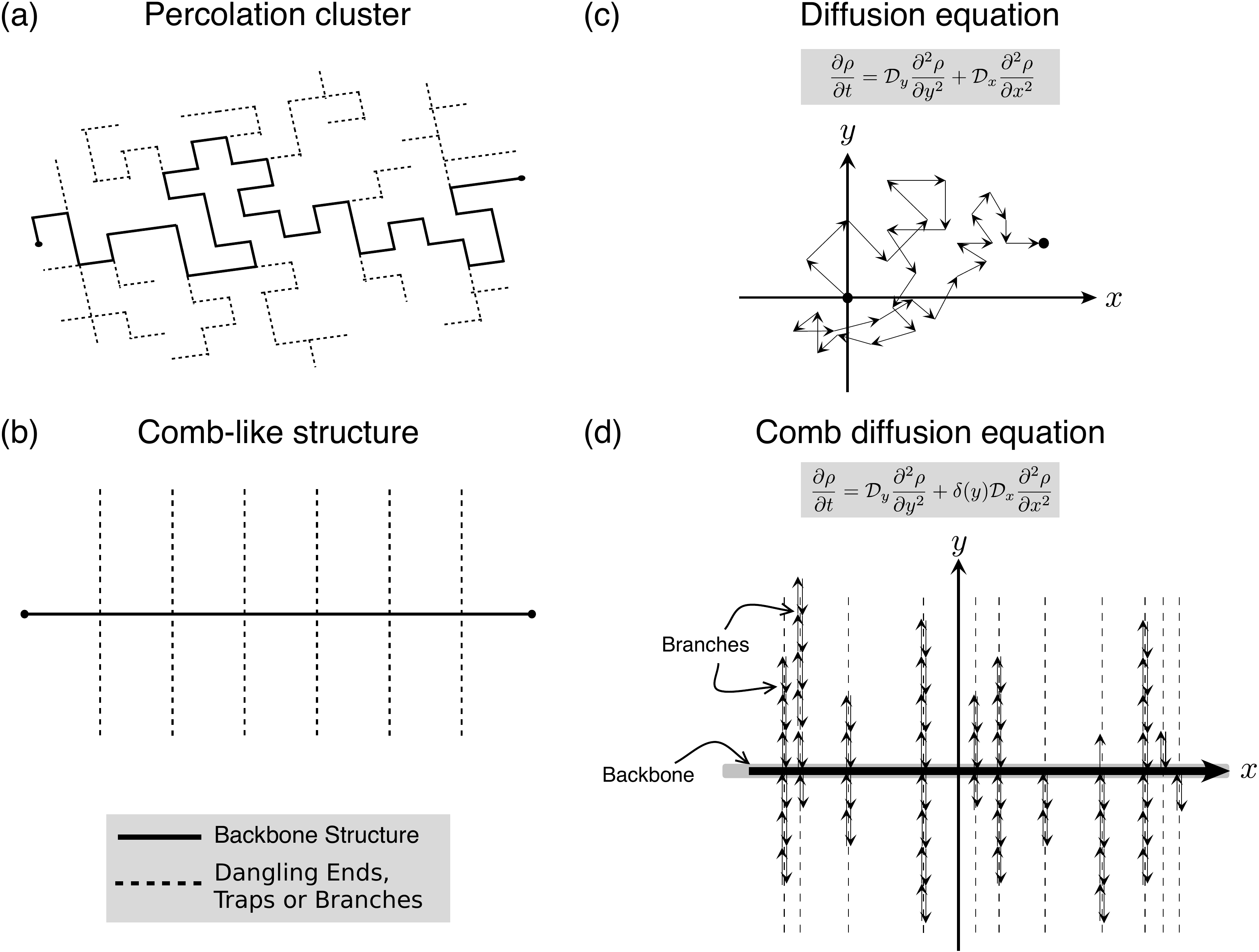}
\caption{The comb-model as simplified description of diffusion on percolation cluster. (a) Illustration of a percolation cluster where the continuous lines indicate the large bond and dashed lines are the dangling ends. The comb-model shown in (b) and modeled by Eq.~(\ref{Comb0}) is a simplified description of the geometry in percolation clusters. Panels (c) and (d) compare the usual two-dimensional diffusive process with the diffusion on comb structure.  The two-dimensional diffusion imposes no geometrical restriction on random walkers, while horizontal shifts occur only when $y=0$ in the comb-model; to accesses different branches, walkers must return to the backbone. Subdiffusive behavior in the backbone is a direct consequence of branches acting like traps.}
\label{fig:1}
\end{figure}

Another model of anomalous diffusion of particular importance to the present work is the comb-model; a model emerged from studies of percolation threshold and anomalous diffusion on fractal structures~\cite{Coniglio1981,Stanley1984,White1984,Havlin1987}. This model describes a diffusive process on a comb-like structure consisting of a ``backbone'' (a single infinite line in the $x$-direction) and ``branches'' (parallel lines in the $y$-direction that intersect the $x$-axis). The comb-model is a simplified description of the fractal geometry of percolation clusters, where the backbone represents the large bond and branches are the remaining bonds or ``dangling ends'' of percolation clusters (Figures~\ref{fig:1}a and \ref{fig:1}b). The comb-model retains essential properties of diffusion on fractals, with the advantage of providing exact results on such complicated systems. Moreover, random walks on comb-like structures established the sojourn times of walkers in the teeth as the underlying mechanism of anomalous diffusion in the backbone.

Given the previously-mentioned experimental findings and because the quenched disorder is intrinsic to the comb structure, it is essential to account for annealed disorder in the comb-model. Here we propose such hybrid comb-models by generalizing the usual comb diffusion equation~\cite{ArBa} via different fractional time-derivative operators. In our hybrid comb-models, different comb-like structures describe quenched disorder, and fractional operators account for annealed disorder. By exploring different configurations where fractional operators act on the branches, backbone, or simultaneously on both, and also by replacing the usual comb structure by a generalized fractal structure, we find a series of nontrivial results that are useful for describing some recent empirical results reported for anomalous diffusion. Among other findings, we observe that these generalized comb-models describe restricted diffusion, Brownian diffusion, and crossovers from subdiffusion to restricted or Brownian diffusions.

The rest of this manuscript is organized as follows. In Section~\ref{sec:2}, we define our generalized version of the comb-model and investigate its solutions under different situations. In Section~\ref{sec:3}, we consider a fractal grid in place the single backbone structure and explore the effects of this modification on the diffusive behavior. Finally, we conclude this work in Section~\ref{sec:4} with a discussion and summary of our findings.

\section{Generalized Comb-Models with Fractional Operators}\label{sec:2}

The diffusion equation for a comb structure was proposed by Arkhincheev and Baskin~\cite{ArBa} and represents a two-dimensional Einstein's diffusion equation where the diffusive term in the $x$-direction is multiplied by a Dirac delta function $\delta(y)$, that is,
\begin{eqnarray}
\label{Comb0}
\frac{\partial}{\partial t}\rho(x,y;t)= 
{\cal{D}}_{y} \frac{\partial^2}{\partial y^2} \rho(x,y;t)+
\delta(y){\cal{D}}_{x} \frac{\partial^2}{\partial x^2} \rho(x,y;t).
\end{eqnarray}
Because of the delta function, diffusion in the $x$-direction only occurs over the backbone structure (when $y=0$). The diffusion in the $y$-direction creates the branch structures; a walker can only leave a branch or access other branches by returning to the backbone structure (Figures~\ref{fig:1}c and \ref{fig:1}d). The geometrical restrictions in Eq.~(\ref{Comb0}) mimic all features of early comb-models, including subdiffusive behavior in the backbone. The trapping times over the branches are also equivalent to a power-law behavior in the waiting-time distributions of a CTRW. The solutions of Eq.~(\ref{Comb0}) are related to a time-fractional diffusion equation (with an anomalous exponent $\alpha_{x}=1/2$) describing the spreading behavior in the backbone~\cite{ArBa,Arkhincheev1999,Wakil2002,IominPRE2005}. The diffusion over the backbone is also described by a time-fractional diffusion equation with exponent $\alpha_{x}=1/4$ in a three-dimensional comb structure and $\alpha_{x}=1/2N$ for an $N$-dimensional case~\cite{Arkhincheev1999}. Extensions of Eq.~(\ref{Comb0}) have been used to obtain a fractional diffusion equation with an absorbent term and a linear external force~\cite{Zahran2009} as well as to deal with generalized fractal structures in the backbone and branches (namely the fractal comb-model)~\cite{Iomin2011,pre2015,jpa2016,sandev2017anomalous}.

In this context, we propose to generalize the comb-model by including different fractional time-derivative operators on the diffusion terms, that is,
\begin{eqnarray}
\label{Comb1}
\frac{\partial}{\partial t}\rho(x,y;t)= 
{\cal{F}}_{t,y}\left({\cal{D}}_{y} \frac{\partial^2}{\partial y^2} \rho(x,y;t)\right) +
\delta(y){\cal{F}}_{t,x}\left({\cal{D}}_{x} \frac{\partial^2}{\partial x^2} \rho(x,y;t)\right)\,,
\end{eqnarray}
where ${\cal{F}}_{t,i}\{ \cdots \}$ is an operator defined by the time derivative of a convolution integral between a function $f(x,y,t)$ and a memory kernel $\mathcal{K}_{i}(t)$ ($i\in \{x,y\}$), that is,
\begin{eqnarray}
\label{KK}
{\cal{F}}_{t,i} \{ f(x,y;t) \}=
\frac{\partial}{\partial t} \int_{0}^{t}f(x,y;t')\mathcal{K}_{i}(t-t')\,dt' \;.
\end{eqnarray}
The use of fractional derivatives in front of spatial operators is motivated by a possible connection with the linear-response theory~\cite{Sokolov2001}. The memory kernel can also be connected with the waiting-time distribution of CTRW and represents a coarse-grained description of the environment's randomness. Specifically, the kernel of the time-convoluted operator represents a density memory (a property of a collection of trajectories) and not a trajectory memory~\cite{Cakir2007}. A derivation of this integro-differential operator and the physical meaning of the memory kernel are given by Sokolov and Klafter~\cite{Sokolov2005}. {It is worth mentioning that different operators~\cite{sokolov2002solutions,PhysRevE.81.011116,PhysRevE.66.046129} have been used to extend diffusion equations. For instance, the operator $\int_{0}^{t}f(x,y;t')\mathcal{K}_{i}(t-t')\,dt'$ was considered by Sokolov~\cite{sokolov2002solutions} for identifying memory kernels that lead non-negative solutions (safe ones) and those that this condition is not guaranteed (dangerous ones).}

The memory kernels $\mathcal{K}_{i}(t)$ define the integro-differential operators in Eq.~(\ref{Comb1}) and establish a connection with fractional time-derivative operators. Thus, Eq.~(\ref{KK}) represents a unified description for a broad class of situations where either singular or non-singular kernels describe different relaxation processes. Moreover, distinct kernels for the $x$ and $y$ directions yield anisotropic diffusion. Equations~(\ref{Comb1}) and (\ref{KK}) recover the usual comb-model [Eq.~(\ref{Comb0})] when 
{{$\mathcal{K}_{x}(t) = \mathcal{K}_{y}(t)=1$}}. In the usual case, there are no memory effects, and geometrical restrictions of the comb-like structure are the only mechanism tied to the anomalous diffusion~\cite{pre2015,IominBook}.

Different choices for $\mathcal{K}_{x}(t)$ and $\mathcal{K}_{y}(t)$ imply in extending the comb-model to different contexts that combine quenched and annealed disorders. One possibility is to consider power-law functions such as 
\begin{eqnarray}
\label{KK11}
\mathcal{K}_{i}(t)=  
 \frac{t^{\alpha_{i}-1}}{\Gamma\left(\alpha_{i}\right)}, 
\end{eqnarray}
which are directly related to the Riemann-Liouville fractional operator~\cite{Livro1} for $0<\alpha_{i}<1$. This fractional operator has been used to investigate several physical contexts, in particular the ones related to anomalous diffusion~\cite{Metzler2000,Barkai2,Book2018}. 

Another possibility is to assume an exponential behavior for the kernels
\begin{eqnarray}\label{KK22e}
\mathcal{K}_{i}(t)&=& {\cal{R}}(\alpha'_{i})\exp\left(-\alpha'_{i} \, t\right)\,,
\end{eqnarray}
where ${\cal{R}}(\alpha'_{i})$ is a normalization constant. {This choice corresponds to the Caputo-Fabrizio operator with $\alpha'_{i}=\alpha_{i}/(1-\alpha_{i})$ \cite{nonsingular1,hristov2017derivatives,tateishifp}}. A remarkable feature of this exponential kernel is its connection with resetting processes~\cite{tateishifp}. In particular, by combining Eqs.~(\ref{Comb1}),~(\ref{KK}), and~(\ref{KK22e}), we find
\begin{eqnarray}
\label{Comb2}
\frac{\partial}{\partial t}\rho(x,y;t)= 
{\cal{D}}_{y} \frac{\partial^2}{\partial y^2} \rho(x,y;t) +
\delta(y){\cal{D}}_{x} \frac{\partial^2}{\partial x^2} \rho(x,y;t)-\tilde{\alpha}\left(\rho(x,y,t)-\varphi(x,y)\right)\,,
\end{eqnarray}
where $\alpha_{x}=\alpha_{y}=\tilde{\alpha}$ and $\varphi(x,y)$ is the initial condition. Equation~(\ref{Comb2}) extends the standard expressions used to analyze resetting processes by including a geometric constraint between the $x$ and $y$ directions. {It is worth noticing that an exponential kernel leads to the Cattaneo equation in the approach of Sokolov~\cite{sokolov2002solutions}, that is, a diffusion-wave equation different from Eq.~(\ref{Comb2}).} The kernel ${\cal K}_{\alpha_{i}}(t)\propto \,E_{\alpha}\left(-\overline{\alpha}t^{\alpha}\right)$, where 
\begin{equation}
E_{\alpha}(z)=\sum_{k=0}^{\infty}\frac{z^k}{\Gamma(\alpha k+1)}  
\end{equation}
is the Mittag-Leffler function~\cite{Livro1} with parameter $\alpha$ and $\overline{\alpha}$ a constant, somehow interpolates between the power-law and exponential cases and has been recently associated with fractional-time derivatives of distributed order~\cite{tateishifp}. It is worth mentioning that these non-singular kernels have been used to investigate different contexts such as diffusion~\cite{tateishifp}, heat processes~\cite{EBasa1}, groundwater flow~\cite{AAtangana1}, and electrical circuits~\cite{JFGomez-Aguilar1}. 

We now focus on the solutions of Eq.~(\ref{Comb1}) in the Fourier-Laplace domain by using the Green function approach. After, we analyze particular cases related to the previous kernels. We consider Eq.~(\ref{Comb1}) subjected to the initial condition $\rho(x,y,0)= \varphi(x,y)$, where $\varphi(x,y)$ is a normalized function, that is, $ \int_{-\infty}^{\infty}dx\int_{-\infty}^{\infty}dy\,\varphi(x,y)=1$. We further assume $\rho(\pm \infty,y;t)=0$ and $\rho(x,\pm \infty;t) = 0$ as boundary conditions. These unlimited boundary conditions avoid possible effects of confinement in a limited domain, making more explicit the impact of geometrical restrictions and fractional operators on the spreading behavior. 

To obtain the solutions of Eq.~(\ref{Comb1}), we first apply the Laplace transform $\left({\mathcal{L}}\left\{\rho(x,y;t)\right\}\right.=$ $\left.\int_{0}^{\infty}e^{-st}\rho(x,y;t)\,dt =\bar{\rho}(x,y;s) \right)$, yielding 
\begin{eqnarray}\label{CombLaplace}
s\bar{\rho}(x,y;s)-\varphi(x,y) = s\bar{\mathcal{D}}_{y}(s)\frac{\partial^2}{\partial y^2} \bar{\rho}(x,y;s)+\delta(y)s\bar{{\cal{D}}}_{x}(s)\frac{\partial^2}{\partial x^2} \bar{\rho}(x,y;s),
\end{eqnarray}
where $\bar{\mathcal{D}}_{y}(s) = {\cal{D}}_{y}\bar{\mathcal{K}}_{y}(s)$ and $\bar{\mathcal{D}}_{x}(s) = {\cal{D}}_{x}\bar{\mathcal{K}}_{x}(s)$. We next apply the Fourier transform $\left({\mathcal{F}}\left\{\rho(x,y;t)\right\}=\tilde{\rho}(k_x,y;s)=\right.$ $\left.\int_{-\infty}^{\infty}e^{-ik_{x}x}\rho(x,y;t)\,dx  \right.)$ on the $x$ variable of  Eq.~(\ref{CombLaplace}), yielding
\begin{eqnarray}
\label{Equation1}
s\bar{{\cal{D}}}_{y}(s)\frac{d^{2}}{dy^{2}}\bar{\rho}(k_{x},y;s) - 
s\left(1-\delta(y) \bar{{\cal{D}}}_{x}(s) k_{x}^{2}\right)\bar{\rho}(k_{x},y;s)  = -\varphi(k_{x},y)\;.
\end{eqnarray}
By using the Green functions approach, the solution for Eq.~(\ref{Equation1}) is written as
\begin{eqnarray}
\label{SolutionLplacerho}
\bar{\rho}(k_{x},y;s) = - \int_{-\infty}^{\infty}dy'\,\varphi(k_{x},y')\bar{{\cal{G}}}(k_{x},y,y';s)\,,
\end{eqnarray}
where the Green function ${\cal{G}}(k_{x},y,y';s)$ is the solution of
\begin{eqnarray}
\label{Green0}
s\bar{{\cal{D}}}_{y}(s)\frac{d^{2}}{dy^{2}}\bar{{\cal{G}}}(k_{x},y,y';s) -\left(s +
\delta(y) s\bar{{\cal{D}}}_{x}(s) k_{x}^{2}\right)\bar{{\cal{G}}}(k_{x},y,y';s)= \delta(y-y')
\end{eqnarray}
subjected to the condition $\bar{{\cal{G}}}(k_{x}, \pm\infty,\bar{y};s)=0$. 

After some calculations, we can show that the solution for Eq.~(\ref{Green0}) is
\begin{eqnarray}
\tilde{\bar{{\cal{G}}}}(k_{x},y,y';s) =
- \frac{1}{2s\sqrt{\bar{{\cal{D}}}_{y}(s)}}e^{-\frac{1}{\sqrt{\bar{{\cal{D}}}_{y}(s)}}|y-y'|} -
\frac{\bar{{\cal{D}}}_{x}(s)k_{x}^{2}}{2\sqrt{\bar{{\cal{D}}}_{y}(s)}}e^{-\frac{1}{\sqrt{\bar{{\cal{D}}}_{y}(s)}}|y|} \bar{{\cal{G}}}(k_{x},0,y';s),
\end{eqnarray}
where
\begin{eqnarray}
\label{BackboneGreen}
\!\!\!\!\!\!\tilde{\bar{{\cal{G}}}}(k_{x},0,y';s) = \frac{e^{-\frac{1}{\sqrt{{\cal{D}}_{y}(s)}}|y'|} }{ s\left({\cal{D}}_{x}(s)k_{x}^{2} +2\sqrt{{\cal{D}}_{y}(s)}\right)}\;
\end{eqnarray}
represents the propagator for the backbone structure (at $y=0$). The term ${\cal{D}}_{y}(s)$ in Eq.~(\ref{BackboneGreen}) indicates that the backbone diffusion explicitly depends on the diffusion occurring along the branches; in other words, memory effects on branches directly affect the diffusion on the backbone.  

The Green function related to Eq.~(\ref{Green0}) subjected to previous boundary condition is thus given by
\begin{eqnarray}
\!\!\!\!\!\!\bar{{\cal{G}}}(k_{x},y,y';s) &=& 
- \frac{1}{2s\sqrt{\bar{{\cal{D}}}_{y}(s)}}e^{-\frac{1}{\sqrt{\bar{{\cal{D}}}_{y}(s)}}|y-y'|} +
\frac{\bar{{\cal{D}}}_{x}(s)k_{x}^{2}}{ \bar{{\cal{D}}}_{x}(s)k_{x}^{2}+ 2\sqrt{\bar{{\cal{D}}}_{y}(s)}} \frac{1}{2s\sqrt{\bar{{\cal{D}}}_{y}(s)}}e^{-\frac{1}{\sqrt{\bar{{\cal{D}}}_{y}(s)}}\left(|y|+|y'|\right)} \nonumber\\
&=& -\frac{1}{2s\sqrt{\bar{{\cal{D}}}_{y}(s)}}\left(
e^{-\frac{1}{\sqrt{\bar{{\cal{D}}}_{y}(s)}}|y-y'|} -  e^{-\frac{1}{\sqrt{\bar{{\cal{D}}}_{y}(s)}}\left(|y|+|y'|\right)} \right) \nonumber \\  &-&\frac{1}{s\left(2\sqrt{\bar{{\cal{D}}}_{y}(s)} + \bar{{\cal{D}}}_{x}(s)k_{x}^{2} \right)}e^{-\frac{1}{\sqrt{\bar{{\cal{D}}}_{y}(s)}}\left(|y|+|y'|\right)} \;.
\end{eqnarray}
After performing the inverse Fourier transform on $x$-direction $( {\mathcal{F}}^{-1}\left\{\tilde{\rho}(k_x,y;t)\right\}$ $=$ $\rho(x,y;t)$ $=$ $\int_{-\infty}^{\infty}e^{ik_{x}x}\tilde{\rho}(k_x,y;t)\,dk_x)$, we obtain
\begin{eqnarray}
\label{Green1}
{\cal{G}}(x,y,y';s) &=& 
-\frac{\delta(x)}{2s\sqrt{\bar{{\cal{D}}}_{y}(s)}} \left( e^{-\frac{|y-y'|}{\sqrt{\bar{{\cal{D}}}_{y}(s)}}} -  e^{-\frac{1}{\sqrt{\bar{{\cal{D}}}_{y}(s)}}\left(|y|+|y'|\right)} \right)
\nonumber \\ &-&
\frac{ 1}{2s\sqrt{2\bar{{\cal{D}}}_{x}(s)\sqrt{\bar{{\cal{D}}}_{y}(s)}}}
e^{-\sqrt{\frac{2\sqrt{\bar{{\cal{D}}}_{y}(s) }}{\bar{{\cal{D}}}_{x}(s)}} |x|}e^{-\frac{1}{\sqrt{{\cal{D}}_{y}(s)}}\left(|y|+y'|\right)}\;.
\end{eqnarray}

\begin{figure}[!t]
\centering
\includegraphics[width=0.7\textwidth]{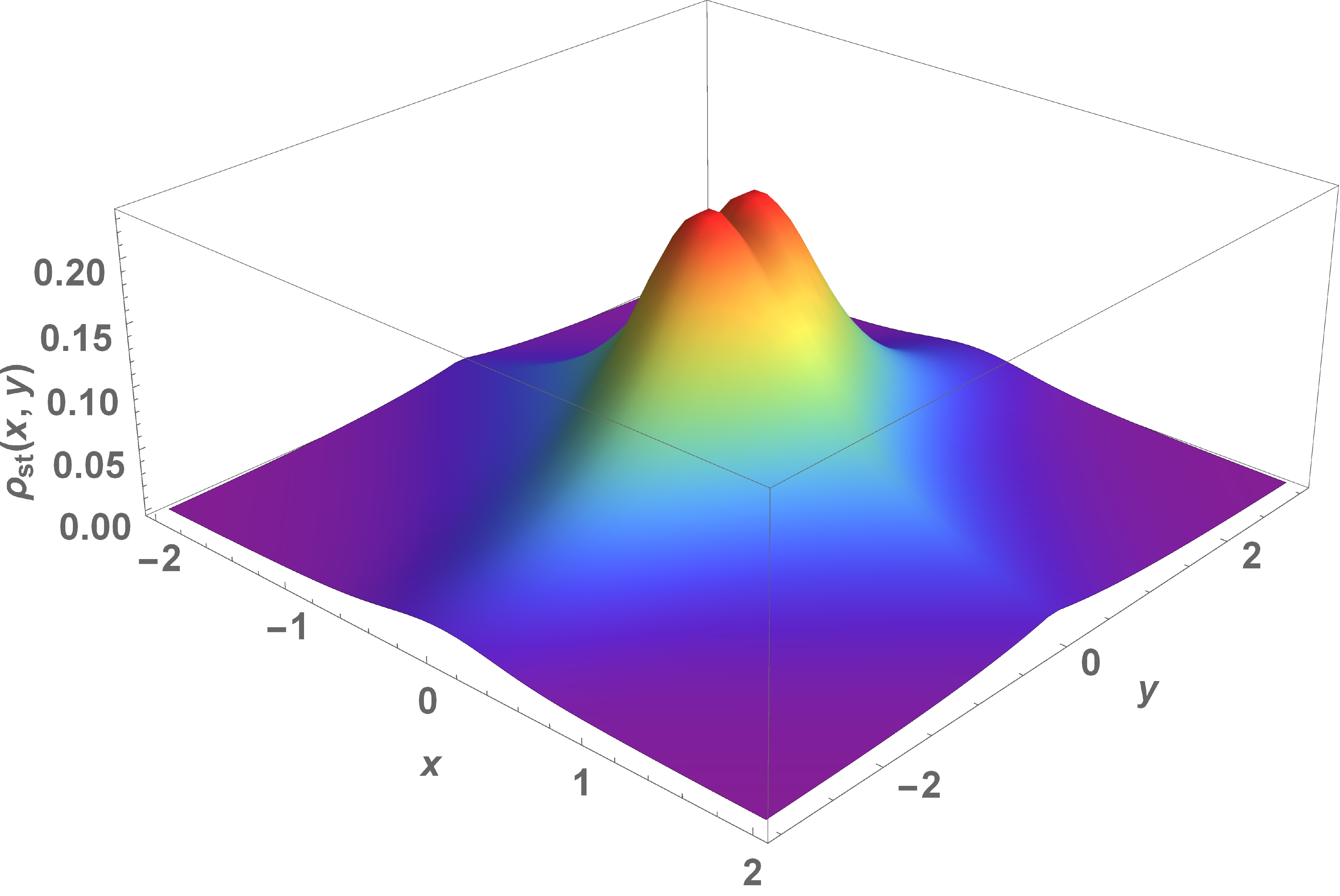}
\caption{Behavior of the stationary distribution obtained from Eq.~(\ref{SolutionIndependentTrho}) with Green function given by Eq.~(\ref{CombModelResetting_Stationary}). For simplicity, we consider the initial condition $\varphi(x,y)\propto e^{-x^{2}/\sigma_{x}^2-y^{2}/\sigma_{y}^2}/\left(\sigma_{x}\sigma_{y}\right)$, where $\sigma_{x}=\sigma_{y}=1/5$), $\sqrt{\alpha/D_{y}}=1$, and $\sqrt{\alpha/D_{x}}=1$ (in arbitrary units). }
\label{fig:2}
\end{figure}

The result in Eq.~(\ref{Green1}) is completely general and can be used to describe different diffusive processes depending on the kernel of the integro-differential operator. For example, for $\bar{{\cal{K}}}_{x}(s)=1/s^{\alpha_{x}}$ and $\bar{{\cal{K}}}_{y}(s)=1/s^{\alpha_{y}}$, the inverse Laplace transform of Eq.~(\ref{Green1}) is
\begin{eqnarray}
\label{GreenRL}
&&{\cal{G}}(x,y,y';t)=- \frac{\delta(x)}{2\sqrt{{\cal{D}}_{y}t^{\alpha_{y}}}}\left\{
{{\Large{H}}}_{1,1}^{1,0} \left[ \frac{|y-y'|}{\sqrt{{\cal{D}}_{y}t^{\alpha_{y}}}}
\left|_{\left(0, 1 \right) }^{\left(1-\frac{\alpha_{y}}{2},  \frac{\alpha_{y}}{2} \right)} \right. \right]-
{{\Large{H}}}_{1,1}^{1,0} \left[ \frac{|y|+|y'|}{\sqrt{{\cal{D}}_{y}t^{\alpha_{y}}}}
\left|_{\left(0, 1 \right) }^{\left(1-\frac{\alpha_{y}}{2},  \frac{\alpha_{y}}{2} \right)} \right. \right]\right\}\nonumber \\ &-& \frac{ 1}{2\sqrt{2{\cal{D}}_{x}\sqrt{{{\cal{D}}}_{y}}}}
\int_{0}^{t}\frac{dt'}{(t-t')t'^{\alpha_{+
}}}{{\Large{H}}}_{1,1}^{1,0} \left[ \sqrt{\frac{2\sqrt{{\cal{D}}_{y}t^{\alpha_{y}}}}{{\cal{D}}_{x}t^{\alpha_{x}}}}|x|
\left|_{\left(0, 1 \right) }^{\left(1-\alpha_{+}, \alpha_{-}\right)} \right. \right]{{\Large{H}}}_{1,1}^{1,0} \left[ \frac{|y|+|y'|}{\sqrt{{\cal{D}}_{y}t^{\gamma_{y}}}}
\left|_{\left(0, 1 \right) }^{\left(0, \frac{\alpha_{y}}{2} \right)} \right. \right]\;,\nonumber\\
\end{eqnarray}
where $\alpha_{+}=\alpha_{x}/2+\alpha_{y}/4$, $\alpha_{-}=\alpha_{x}/2-\alpha_{y}/4$, and $H_{p,q}^{m,n}\left[z\left|^{(a_p,A_p)}_{(b_q,B_q)} \right.\right]$ is the Fox H function~\cite{Fox}. The case ${\cal{K}}_{x}(s)=1/(s+ \alpha)$ and ${\cal{K}}_{y}(s)=1/(s+\alpha)$ (where $\alpha'_{x}=\alpha'_{y}=\alpha$) lead us to
\begin{eqnarray}
\label{CombModelResetting}
&&{\cal{G}}(x,y,y';t)=-\delta(x)\int_{0}^{t}dt'\,k(t,t')\frac{e^{-\alpha t'}}{\sqrt{4\pi D_{y}t}}\!\left\{\!e^{-\frac{|y-y'|^2}{4 D_{y}t'}}-e^{-\frac{(|y|+|y'|)^2}{4 D_{y}t'}}\right\}\nonumber \\ &-& \int_{0}^{t}\!\!dt'\,k(t,t')\!\!\int_{0}^{t'}\!\!dt''\,\!\frac{|y|+|y'|}{\sqrt{8D_{x}t''\sqrt{D_{y}t''}}}{{\Large{H}}}_{1,1}^{1,0} \left[ \left(\frac{4D_{y}}{D_{x}^{2}t''}\right)^{\frac{1}{4}}|x|
\left|_{\left(\frac{3}{4},  \frac{1}{4} \right) }^{\left(\frac{1}{4},  \frac{1}{4} \right)} \right. \right]\frac{e^{-\frac{(|y|+|y'|)^{2}}{4D_{y}(t'-t'')}}}{\sqrt{4\pi\left[ D_{y} \left(t'-t''\right)\right]^{3}}}\;,\nonumber\\
\end{eqnarray}
with $k(t,t')=\alpha+\delta(t-t')$ and $D_{x(y)}={\cal{D}}_{x(y)}{\cal{R}}(\alpha_{x(y)})$. Differently from Eq.~(\ref{GreenRL}), Eq.~(\ref{CombModelResetting}) has a stationary solution (a time-independent solution ${\cal{G}}_{st}(x,y,y')=\lim_{t\rightarrow\infty}{\cal{G}}(x,y,y';t)$) given by
\begin{eqnarray}
\label{CombModelResetting_Stationary}
{\cal{G}}_{st}(x,y,y') &=&    -
\sqrt{\frac{\alpha}{4{D}_{y}}} \delta(x)\left( e^{-\sqrt{\frac{\alpha}{D_{y}}}|y-y'|} -  e^{-\sqrt{\frac{\alpha}{D_{y}}}\left(|y|+|y'|\right)} \right)
\nonumber \\ &-&
\sqrt{\frac{ \alpha\sqrt{\alpha}}{8D_{x}\sqrt{D_{y}}}}
e^{-\sqrt{\frac{2\alpha}{D_{x}}\sqrt{\frac{D_{y} }{\alpha}}} |x|}e^{-\sqrt{\frac{\alpha}{{D}_{y}}}\left(|y|+|y'|\right)}\;.
\end{eqnarray}
This result is obtained from an arbitrary condition where
\begin{eqnarray}
\lim_{s\rightarrow 0}\bar{{\cal{K}}}_{x}(s)=const \;\;\;{\text{and}}\;\;\;
\lim_{s\rightarrow 0}\bar{{\cal{K}}}_{y}(s)=const, \nonumber
\end{eqnarray}
which implies in $\lim_{s\rightarrow 0}\left(s\bar{{\cal{G}}}(x,y,y';s)\right)={\cal{G}}_{st}(x,y,y')$. Figure~\ref{fig:2} illustrates the behavior of Eq.~(\ref{SolutionLplacerho}) when considering the Green function of Eq.~(\ref{CombModelResetting_Stationary}). In this stationary limit, the solution in Eq.~(\ref{SolutionLplacerho}) is
\begin{eqnarray}
\label{SolutionIndependentTrho}
\rho_{st}(x,y) = \int_{-\infty}^{\infty}dx'\int_{-\infty}^{\infty}dy'\,\varphi(x',y'){\cal{G}}_{st}(x-x',y,y')\;.
\end{eqnarray}

The MSD also carries information about the medium structure. We thus use the previous results to investigate how the MSD in each direction changes under those different scenarios. To do so and avoid transient behaviors related to the initial position of the walkers, we consider the initial condition $\varphi(x,y)=\delta(x)\delta(y)$. Under these assumptions, the MSD in the Laplace domain for each direction is
\begin{eqnarray}
\label{msdy}
\bar{\sigma}_{y}^{2}(s)=\left\langle \left(y- \langle y \rangle\right)^{2}\right\rangle &=&  2\bar{{\cal{D}}}_{y}(s)/s \;\;\; {\text{and}}\\ 
\label{msdx}
\bar{\sigma}_{x}^{2}(s)=\left\langle \left(x- \langle x \rangle \right)^{2} \right \rangle &=&  \bar{{\cal{D}}}_{x}(s)/\left(s\sqrt{\bar{{\cal{D}}}_{y}(s)}\right).
\end{eqnarray}
Equations~(\ref{msdy}) and (\ref{msdx}) show that the MSD in the $y$-direction depends only on its memory kernel, while the MSD in the $x$-direction depends on both memory kernels. These features naturally emerge in the time-domain; indeed, by performing the inverse Laplace transform, we find
\begin{eqnarray}
\sigma_{y}^{2}(t)=2\int_{0}^{t}dt'\,{\cal{D}}_{y}(t') \;\;\; {\text{and}}\;\;\; 
\sigma_{x}^{2}(t) =  
\frac{{\cal{D}}_{x}}{\sqrt{{\cal{D}}_{y}}}\int_{0}^{t}dt'\,\zeta_{x,y}(t') \nonumber
\end{eqnarray}
where
\begin{equation}
\zeta_{x,y}(t)=\mathcal{L}^{-1}\left\{\frac{\bar{\mathcal{K}}_{x}(s)}{\sqrt{\bar{\mathcal{K}}_{y}(s)}}\right\}.    
\end{equation}
These dependencies are a direct consequence of the comb structure and are somehow related to the results of Ref.~\cite{Ribeiro2014}. The authors of that work have simulated fractional Brownian walks on a comb-like structure and reported that memory effects (associated with Hurst exponents) in $x$-direction do not affect the diffusive behavior in the $y$-direction; but, in the backbone, they found a nontrivial interplay between long-range memories in $x$ and $y$-directions.

We now consider the behavior of the system in each direction. To do so, we note that the Green function for the probability distribution function along the backbone $\mathcal{G}_{1}(x,t)=\int_{-\infty}^{\infty}dy\,\rho(x,y,t)$ satisfies the following generalized diffusion equation
\begin{eqnarray}\label{pdf x}
\frac{\partial}{\partial t}\mathcal{G}_{1}(x,t)=\frac{\mathcal{D}_{x}}{2\sqrt{\mathcal{D}_{y}}}\frac{\partial}{\partial t}\int_{0}^{t}dt'\,\zeta_{x,y}(t-t')\frac{\partial^{2}}{\partial x^{2}}\mathcal{G}_{1}(x,t')\;.
\end{eqnarray}
Similarly, the corresponding generalized diffusion equation for the Green function along the branches $\mathcal{G}_{2}(y,t)=\int_{-\infty}^{\infty}dx\,\rho(x,y,t)$ is
\begin{eqnarray}\label{pdf y}
\frac{\partial}{\partial t}\mathcal{G}_{2}(y,t)=\mathcal{D}_{y}\frac{\partial}{\partial t}\int_{0}^{t}dt'\,\mathcal{K}_{y}(t-t')\frac{\partial^{2}}{\partial y^{2}}\mathcal{G}_{2}(y,t').
\end{eqnarray}
These two forms suggest that both Eqs.~(\ref{pdf x}) and (\ref{pdf y}) have similar mathematical properties. By following Refs.~\cite{csf2017,fcaa2018}, we can verify that the probability distribution functions $\mathcal{G}_{1}(x,t)$ and $\mathcal{G}_{2}(y,t)$ are non-negative if: $1/\left[s\bar{\zeta}_{x,y}(s)\right]$  and $1/\left[s\bar{\mathcal{K}}_{y}(s)\right]$ are completely monotone functions; and $1/\bar{\zeta}_{x,y}(s)$ and $1/\bar{\mathcal{K}}_{y}(s)$ are Bernstein functions. Moreover, by following the results of Ref.~\cite{Sokolov2001}, we can further verify that Eqs.~(\ref{pdf x}) and (\ref{pdf y}) fulfill the Nyquist theorem, and consequently, their solutions are thermodynamically sound.

To investigate the interplay between mechanisms of annealed (memory kernels) and quenched (comb-like structure) disorders, we start considering several definitions for the fractional time operators. For the sake of comparison, it is worth remembering that ordinary derivatives (usual comb-model) imply in $\mathcal{K}_{x}(t) = \mathcal{K}_{y}(t)=\delta(t)$, that in turn lead to $\sigma_{y}^{2}(t)\propto t$ and $\sigma_{x}^{2}(t)\propto t^{1/2}$~\cite{ArBa}.

We first consider the Riemann-Liouville operator, yielding the kernels $\bar{{\cal{K}}}_{\alpha_{x}}(s)\propto 1/s^{\alpha_{x}}$ ($0<\alpha_{x}<1$) and $\bar{{\cal{K}}}_{\alpha_{y}}(s)\propto 1/s^{\alpha_{y}}$ ($0<\alpha_{y}<1$). These memory kernels are related to the Green function given by Eq.~(\ref{GreenRL}), and their corresponding MSDs are
\begin{eqnarray}
\label{y2RL}
\sigma_{y}^{2}(t)&=&  2\mathcal{D}_{y}\,\mathcal{L}^{-1}\left\{s^{-\alpha_{y}-1}\right\}=2\mathcal{D}_{y}\frac{t^{\alpha_{y}}}{\Gamma(1+\alpha_{y})} \;\;\;\;\ \text{and} \\ 
\label{x2RL}
\sigma_{x}^{2}(t) &=&  \frac{\mathcal{D}_{x}}{\sqrt{\mathcal{D}_{y}}}\mathcal{L}^{-1}\left\{s^{-\alpha_{x}+\alpha_{y}/2-1}\right\}=\frac{\mathcal{D}_{x}}{\sqrt{\mathcal{D}_{y}}}\frac{t^{\alpha_{x}-\alpha_{y}/2}}{\Gamma(1+\alpha_{x}-\alpha_{y}/2)}.
\end{eqnarray}
Equation~(\ref{y2RL}) shows that the diffusion in branches is independent of the backbone dynamics; it only depends on its memory effects. However, the MSD in the backbone depends on memory effects in both directions, as shown in Eq.~(\ref{x2RL}).

\begin{figure}[!t]
\centering
\includegraphics[width=0.6\textwidth]{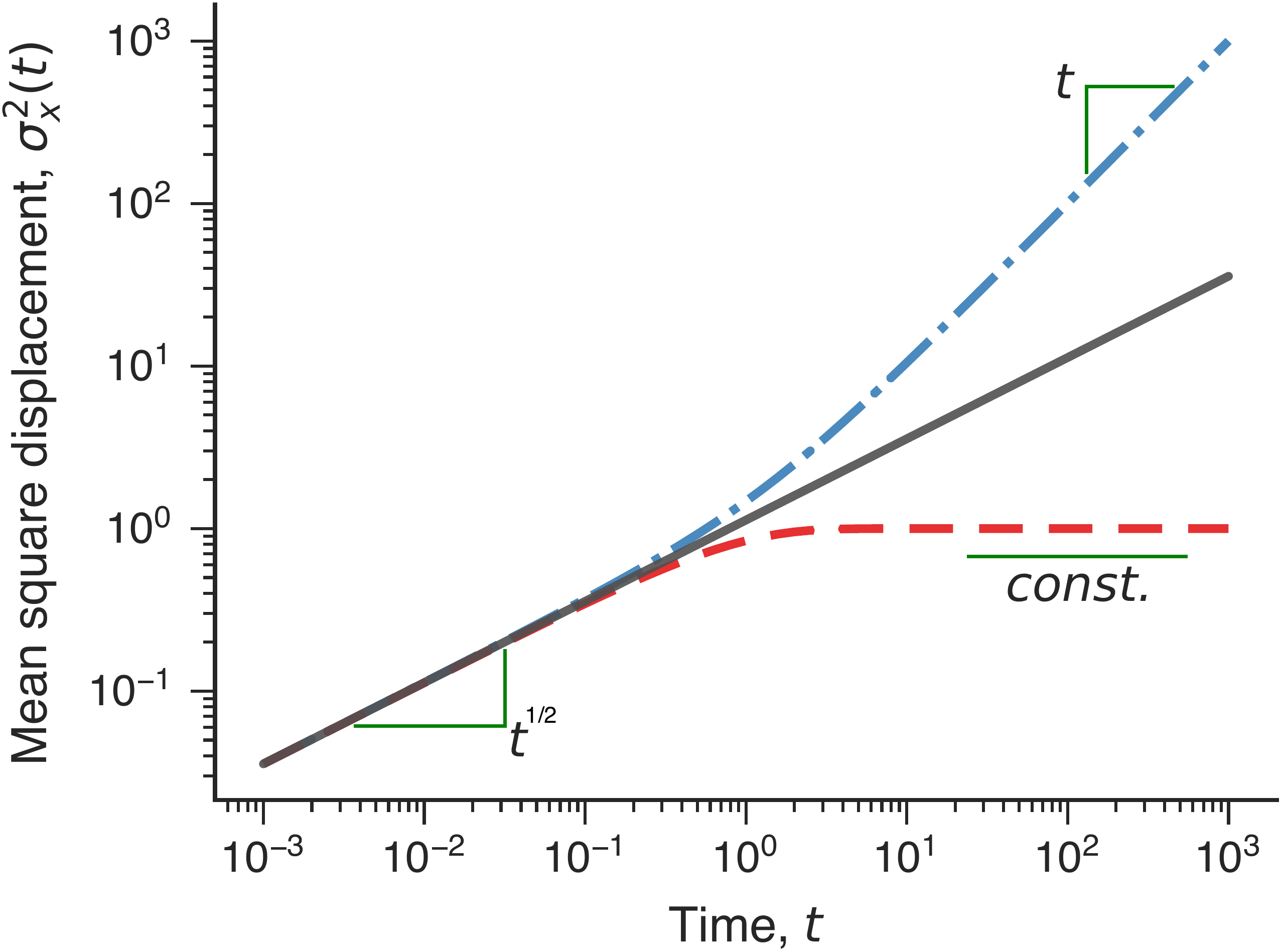}
\caption{Behavior of the MSD $\sigma_{x}^{2}$ versus $t$ for different kernels. The solid black line refers to subdiffusion when $\bar{{\cal{K}}}_{x}(s)\propto 1/s^{\alpha_{x}}$ and $\bar{{\cal{K}}}_{y}(s)\propto 1/s^{\alpha_{y}}$, where we have chosen ${\cal{D}}_{x}/\sqrt{{\cal{D}}_{y}}=1$, $\alpha_{x}=1$, and $\alpha_{y}=1$. The dot-dashed blue line corresponds to usual diffusion obtained as the asymptotic behavior for $\bar{{\cal{K}}}_{x}(s)\propto 1/s^{\alpha_{x}}$ and $\bar{{\cal{K}}}_{y}(s)\propto 1/\left(s+\alpha'_{y}\right)$, where, for simplicity, we have chosen ${\cal{D}}_{x}/\sqrt{{\cal{D}}_{y}{\cal{R}}(\alpha_{y})}=1$, $\alpha_{x}=1$, and $\alpha_{y}=1$.  The dashed red line corresponds to $\bar{{\cal{K}}}_{x}(s)\propto 1/(s+\alpha_{x})$ with $\bar{{\cal{K}}}_{y}(s)\propto 1/(s+ \alpha_{y})$ (the case with stationary state), where we have chosen ${\cal{R}}(\alpha_{x}){\cal{D}}_{x}/\sqrt{{\cal{D}}_{y}{\cal{R}}(\alpha_{y}})=1$, $\alpha_{x}=1$, and $\alpha_{y}=1$.}
\label{fig:3}
\end{figure}

By imposing the conditions of non-negativity to the corresponding solution, we find that $\alpha_{x}>\alpha_{y}/2$. To better understand this condition, let us examine some limiting cases. When $\alpha_{x}=1$ (ordinary derivative in the backbone) and $0<\alpha_{y}<1$ (memory effects in the branches), the backbone diffusion is enhanced if $1/2< 1-\alpha_{y}/2<1$. This result is intriguing and counter-intuitive because, as the branches act like traps, we could initially presume that the slower the diffusion in the branches, the more subdiffusive is the diffusion on the backbone; but quite the opposite happens. When the spread over the branches is subdiffusive, the walkers stay closer to the backbone and their probability of returning to the backbone increases, enhancing the diffusion in the $x$-direction. This phenomenon is related to the so-called ``subdiffusion paradox'' reported in cell environments~\cite{Sereshki2012,BarkaiPhysTod2012,Barr2015}. Although subdiffusion reduces the exploration area, it increases the likelihood of walkers to stay close to specific targets~\cite{Golding2006,Guigas2008}. In contrast, for $1/2<\alpha_{x}<1$ (memory effects in the backbone) and $\alpha_{y}=1$ (ordinary derivative in the branches), the spread in the backbone is even more subdiffusive if $0< \alpha_{x}-1/2<1/2$. Thus, the backbone subdiffusion is governed by the interplay of two mechanisms: the trapping in the branches and the memory effects in the backbone. Furthermore, an essential feature of Eqs.~(\ref{y2RL}) and (\ref{x2RL}) is scale invariance, that is, the effects of geometrical restrictions and memory effects are the same in all time-scales. This last behavior is illustrated in Figure~\ref{fig:3} for $\alpha_{x}=1$ and $\alpha_{y}=1$ (solid black line).

As a second example, we investigate an anisotropic case characterized by different memory kernels for $x$ and $y$ directions. We maintain the same power-law kernel for the backbone, that is, ${\cal{K}}_{\alpha_{x}}(s)\propto 1/s^{\alpha_{x}}$ {(with $1/2\leq\alpha_{x}<1$ to ensure non-negative solutions)}, and consider an exponential memory kernel ${\cal{K}}_{\alpha_{y}}(s)\propto 1/\left(s+\alpha'_{y}\right)$ for the branches, with $\alpha'_{y}=\alpha_{y}/(1-\alpha_{y})$ and $0<\alpha_{y}<1$. These choices correspond to the Riemann-Liouville fractional operator in the $x$-direction and the Caputo-Fabrizio operator in the $y$-direction. Under these conditions, we find the MSDs
\begin{eqnarray}
\label{y2RLCF}
\sigma_{y}^{2}(t) &=&  2D_{y}\,\mathcal{L}^{-1}\left\{\frac{s^{-1}}{s+\alpha'_{y}}\right\}=2\frac{D_{y}}{\alpha'_{y}}\left(1-e^{-\alpha'_{y}t}\right) \;\;\; \text{and} \;\;\; \\ 
\label{x2RLCF}
\sigma_{x}^{2}(t) &=&  \frac{\mathcal{D}_{x}}{\sqrt{D_{y}}}\mathcal{L}^{-1}\left\{\frac{s^{-\alpha_{x}-1}}{\left(s+\alpha'_{y}\right)^{-1/2}}\right\}=\frac{\mathcal{D}_{x}}{\sqrt{D_{y}}}t^{\alpha_{x}-1/2}E_{1,\alpha_{x}+1/2}^{-1/2}(-\alpha_{y}t),
\end{eqnarray}
where
\begin{equation}\label{ML three}
E_{\alpha,\beta}^{\delta}(z)=\sum_{k=0}^{\infty}\frac{(\delta)_k}{\Gamma(\alpha k+\beta)}\frac{z^k}{k!}
\end{equation}
is the three-parameter Mittag-Leffler function~\cite{Prabhakar}, and $(\delta)_k=\Gamma(\gamma+k)/\Gamma(\gamma)$ represents the Pochhammer symbol~\cite{Prabhakar}. For the calculations of Eqs.~(\ref{y2RLCF}) and (\ref{x2RLCF}) we have used that~\cite{Prabhakar}
\begin{equation}\label{laplaceML}
\mathcal{L}\left\{t^{\beta-1}E_{\alpha,\beta}^{\delta}(-\nu t^{\alpha})\right\}(s)=\frac{s^{\alpha\delta-\beta}}{\left(s^{\alpha}+\nu\right)^{\delta}},
\end{equation}
where $\Re(s)>|\nu|^{1/\alpha}$. 

Equation~(\ref{y2RLCF}) shows the isolated effects of the exponential memory kernel. We notice that the exponential term approaches zero for long times, and the MSD thus reaches a plateau of saturation. This behavior describes a confined (localized, restricted, or corralled) diffusion, where $\alpha'_{y}$ can be associated with a saturation rate, and the asymptotic value of the MSD represents the magnitude of the confinement region. MSDs having the general form of Eq.~(\ref{y2RLCF}), that is, 
\begin{equation}
\label{ConfinedDiffusion}
    \sigma^{2}(t) = A_{\infty}(1-e^{-\xi t}),
\end{equation}
are well-known to emerge in Ornstein-Uhlenbeck processes~\cite{Uhlenbeck1945} and restricted diffusion confined within reflecting boundaries~\cite{kusumi1993}. However, the restricted diffusion observed here occurs without external forces or finite boundary conditions, a remarkable feature of the Caputo-Fabrizio operator that has also been reported in Ref.~\cite{tateishifp}.

The confined diffusion observed in Eq.~(\ref{ConfinedDiffusion}) suggests a relationship between the exponential kernels and stationary states (stochastic localization phenomena). Indeed, the same equation emerges in the work M\'endez and Campos~\cite{Campos2016}, where a CTRW model for diffusion with resetting (walkers return to the origin with a resetting probability $r$) was proposed. A connection between fractional diffusion equations with the Caputo-Fabrizio operator and diffusion with stochastic resetting was also established in Ref.~\cite{tateishifp}. M\'endez and colleagues \cite{Mendez2015} also studied a CTRW on a comb structure subjected to a bias parameter on the branches, where Eq.~(\ref{ConfinedDiffusion}) appears as an asymptotic behavior for the backbone diffusion when the walker is biased to stay along the branches. The authors of Ref.~\cite{Ribeiro2014} verified that normal diffusion emerges on the backbone when a fractional Brownian motion with long-range anti-persistent correlations occurs on the branches. By studying a minimal random walk model with infinite memory (walkers preferentially return to previously visited sites), Boyer and Solis-Salas~\cite{Boyer2014} established a connection between long-range memory and stationary states of MSD and demonstrated how to infer memory strength use in animals (monkeys).

In this context, we can verify whether the relation between the Caputo-Fabrizio operator and confined diffusion is valid for the comb-model from the evolution of Eq.~(\ref{x2RLCF}). To do so, we calculate the asymptotic limits of Eqs.~(\ref{y2RLCF}) and (\ref{x2RLCF}) for short- and long-times, that is,
\begin{eqnarray}
\label{y2limits}
\sigma_{y}^{2}(t) &\sim&  2\mathcal{D}_{y}\,\left\lbrace \begin{array}{c l} 
\alpha'_{y}t, \quad & t\rightarrow0,\\
1, \quad & t\rightarrow\infty,  
\end{array}\right.~~\text{and}\\ 
\label{x2limits}
\sigma_{x}^{2}(t) &\sim&  \frac{\mathcal{D}_{x}}{\sqrt{D_{y}}}\left\lbrace \begin{array}{c l} 
\frac{t^{\alpha_{x}-1/2}}{\Gamma(\alpha_{x}+1/2)}, \quad & t\rightarrow0,\\
\frac{t^{\alpha_{x}}}{\Gamma(\alpha_{x}+1)}, \quad & t\rightarrow\infty.
\end{array}\right.
\end{eqnarray}
In previous calculations, we use the formula~\cite{fcaa2015,GG}
\begin{equation}
E_{\alpha,\beta}^{\delta}(-z)=\frac{z^{-\delta}}{\Gamma(\delta)}\sum_{k=0}^{\infty}\frac{\Gamma(\delta+k)}{\Gamma(\beta-\alpha(\delta+n))}\frac{(-z)^{-n}}{n!},
\end{equation}
for $0<\alpha<2$ and $z\rightarrow\infty$, from which we find the asymptotic behavior 
\begin{equation}
E_{\alpha,\beta}^{\delta}(-t^{\alpha})\simeq\frac{t^{-\alpha\delta}}{\Gamma(\beta-\alpha\delta)}, \quad t\rightarrow\infty.
\end{equation} 
Equations~(\ref{y2limits}) and (\ref{x2limits}) show that Brownian motion governs the branches dynamics at short-times, promoting enhanced subdiffusion in the backbone with $0<\alpha'_{x}-1/2<1/2$ and $\alpha'_{y}=1$ (as discussed earlier). In the long-time limit, a stationary state emerges in the branches, and the backbone dynamics only depends on its power-law memory kernel. In particular, there is a crossover from subdiffusion ($\sigma_{x}^{2}(t) \propto t^{1/2}$) to Brownian diffusion ($\sigma_{x}^{2}(t) \propto t$) when $\alpha_{x}=1$ (dot-dashed blue line in Figure~\ref{fig:3}). We thus find an intriguing result where the interplay between geometrical restriction and memory effects (mechanisms associated with subdiffusion) produces usual Brownian motion. Similar behavior also emerges for power-law memory kernels when $\alpha_{x}\rightarrow 1$ and $\alpha_{y}\rightarrow 0$, for suitable combinations of Hurst exponents in fractional Brownian motions over a comb structure (Figure~5 of Ref.~\cite{Ribeiro2014}), and without memory effects when the branches of the comb are finite~\cite{Havlin2002, Berez2015,Arkhincheev2007}. However, the results of Eqs.~(\ref{y2limits}) and (\ref{x2limits}) are obtained with no memory effects in the $x$-direction and the stationary state is a consequence of the exponential memory kernel valid for $0<\alpha'_{y}<1$.

The crossover is an essential feature of our model and provides insights into the time scale that each mechanism of subdiffusion is most relevant. As we already discussed, the dynamics in short-times is the same as the usual comb-model; therefore, subdiffusion is caused by geometrical restrictions. On the other hand, the exponential memory kernel produces a dynamics similar to a random walk with a high probability of returning to the origin. Since the backbone is at the origin, walkers along the branches tend to stay confined near the backbone because of memory effects, which in turn produces Brownian diffusion on the backbone. The memory effects along the branches thus dominate longer time scales. This interpretation is somehow in agreement with the results on anisotropic diffusion of entangled biofilaments reported in Ref.~\cite{Tsang2017}, where the authors have written: {\it ``The physical reason is that linear macromolecules become transiently localized in directions transverse to their backbone but diffuse with relative ease parallel to it.''} In particular, they obtained an empirical MSD $\sim t^{0.2}$ for the transversal direction and a MSD $\sim t^{0.9}$ for the parallel direction of such macromolecules. Liang Hong and co-workers also reported a gradual crossover from subdiffusion to Brownian diffusion on the mobility of water molecules on protein surfaces~\cite{HongPRL2018}. They further argued that a broad distribution of trapping times causes the subdiffusion; however, water molecules start jumping to the empty sites as the trappings become occupied, resulting in the Brownian diffusion.

We can further investigate the effects of exponential memory kernels simultaneously acting on the backbone and branches, that is, $\bar{{\cal{K}}}_{\alpha_{x}}(s)\propto 1/(s+\alpha'_{x})$ and $\bar{{\cal{K}}}_{\alpha_{y}}(s)\propto 1/(s+ \alpha'_{y})$. These memory kernels are related to the Green functions given by Eqs.~(\ref{CombModelResetting}) and (\ref{CombModelResetting_Stationary}), where $\alpha'_{x}=\alpha'_{y}=\alpha$ {is also a condition ensuring the non-negativity of the corresponding solution}. If our interpretation of the exponential memory kernel is valid, we expected stationary states to emerge even in the backbone dynamics. This hypothesis is corroborated by Figure~\ref{fig:2} and the results for MSD 
\begin{eqnarray}
\sigma_{y}^{2}(t) &=&  2\mathcal{D}_{y}\,\mathcal{L}^{-1}\left\{\frac{s^{-1}}{s+\alpha}\right\}=2\frac{D_{y}}{\alpha}\left(1-e^{-\alpha t}\right) \;\;\; \text{and} \\ 
\sigma_{x}^{2}(t) &=&  \frac{D_{x}}{\sqrt{D_{y}}}\mathcal{L}^{-1}\left\{\frac{\sqrt{s+\alpha}}{s(s+\alpha)}\right\}=\frac{D_{x}}{\sqrt{D_{y}}}\frac{1}{\sqrt{\alpha}}\mathrm{erf}\left(\sqrt{\alpha t}\right),
\end{eqnarray}
where $\mathrm{erf}(x)$ is the error function. These results show that the behavior on both directions reaches a stationary state for long times, that is, $\sigma_{y}^{2}(t)$ and $\sigma_{x}^{2}(t)$ approach a constant plateau when $t \rightarrow \infty$. Figure~\ref{fig:3} shows the behavior of $\sigma_{x}^{2}(t)$ for different time scales (dashed red line). Once again, a crossover characterizes the backbone dynamics and the system evolves from subdiffusion to confined diffusion (stationary state). The quenched mechanism dominates at short-time scales and the annealed mechanism predominates in the long-run. It is noteworthy that confined diffusion on both $x$ and $y$ directions has been experimentally observed in the crowded environment of living cells such as in lateral diffusion of membrane receptors~\cite{kusumi1993} and diffusion of protein aggregates in live {\it E. coli} cells~\cite{Coquel2013}. 

\section{Generalized Fractal Structure of Backbones}\label{sec:3}

\begin{figure}[!t]
\centering
\includegraphics[width=0.5\textwidth]{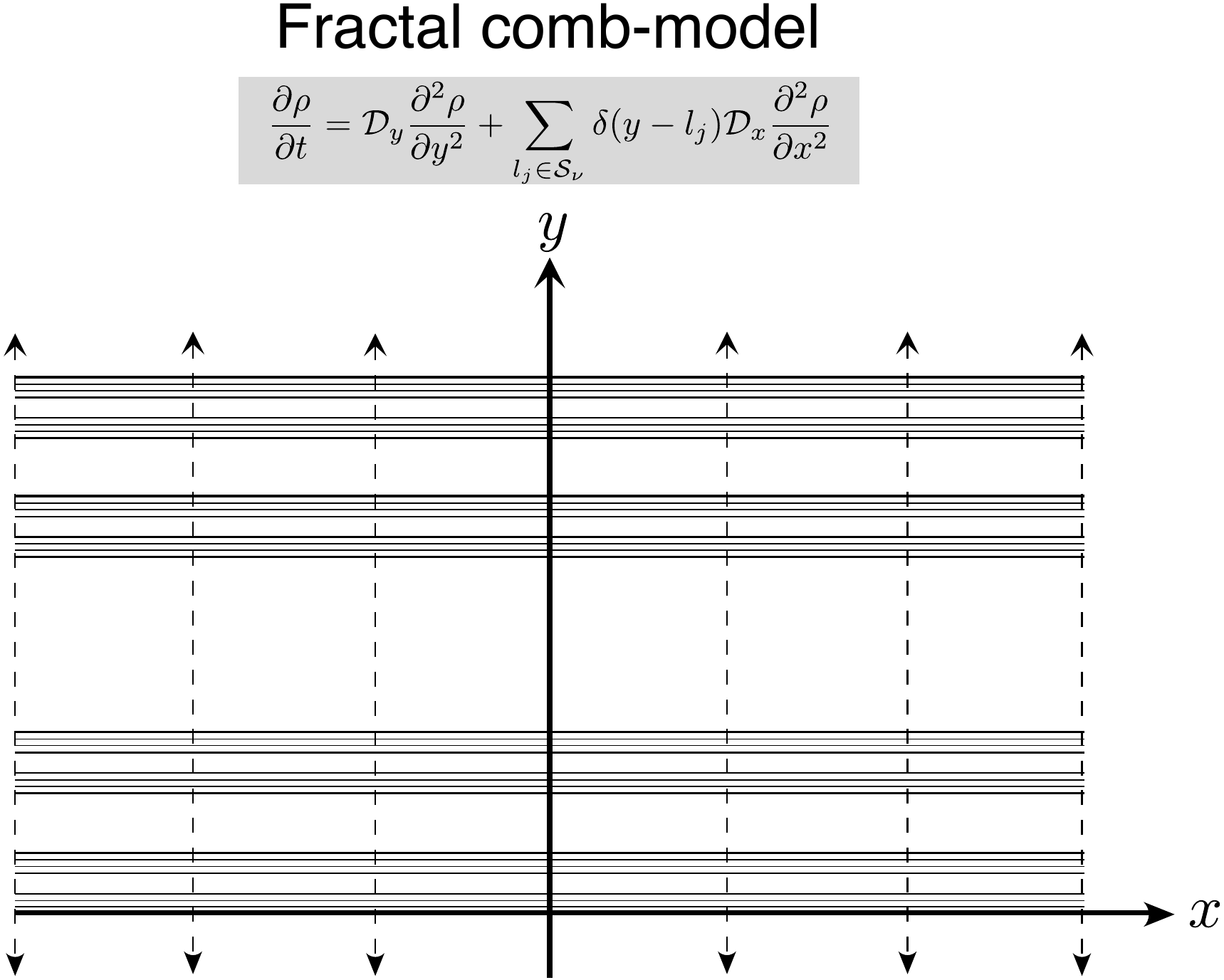}
\caption{Example of a fractal comb structure. The one-third Cantor set (at the third step of construction) provides the rule to locate the backbones perpendicularly to the $y$-axis. The location of the branches is distributed continuously along the backbones, that is, walkers access the branches through any position on the backbones. This spatial configuration characterizes a fractal grid.}
\label{fig4}
\end{figure}

We now focus on generalizing the quenched disorder mechanism of the comb-model (Eq.~\ref{Comb1}) by changing its geometrical restrictions. Instead of multiplying the diffusion term in the $x$-direction by a single delta function $\delta(y)$, we consider a multiplication by $\sum_{l_{j}\in\mathcal{S}_{\nu}}\delta(y-l_{j})$ to obtain an infinite number of backbones, where the position of the backbones $l_{j}$ $(j=1,2,\dots)$ belong to a fractal set $\mathcal{S}_{\nu}$ with fractal dimension $0<\nu<1$. These geometrical restrictions characterize a fractal grid~\cite{pre2015,jpa2016,sandev2017anomalous}, as illustrated in Figure~\ref{fig4} for the one-third Cantor set ($\nu \simeq 0.631$) at the third step of the iteration. Under these conditions, the diffusion equation for the fractal comb-model is
\begin{eqnarray}
\label{Comb1 grid}
\frac{\partial}{\partial t}\rho(x,y;t)= 
{\cal{F}}_{t,y}\left({\cal{D}}_{y} \frac{\partial^2}{\partial y^2} \rho(x,y;t)\right) +
\sum_{l_{j}\in\mathcal{S}_{\nu}}\delta(y-l_{j}){\cal{F}}_{t,x}\left({\cal{D}}_{x} \frac{\partial^2}{\partial x^2} \rho(x,y;t)\right).
\end{eqnarray}

We proceed by defining the generalized diffusion equation governing the dynamics on the backbones. By applying the Laplace transform in Eq.~(\ref{Comb1 grid}), we have
\begin{eqnarray}\label{comb laplace2}
\!\!\!\! s\bar{\rho}(x,y;s)-\varphi(x,y) = s{\cal{D}}_{y}\bar{\mathcal{K}}_{y}(s)\frac{\partial^2}{\partial y^2} \bar{\rho}(x,y;s)+\sum_{l_{j}\in\mathcal{S}_{\nu}}\delta(y-l_{j})s{\cal{D}}_{x}\bar{\mathcal{K}}_{x}(s)\frac{\partial^2}{\partial x^2} \bar{\rho}(x,y;s).
\end{eqnarray}
By following the procedures of Ref.~\cite{jpa2016} and considering a suitable initial condition, we can represent the probability distribution function as
\begin{eqnarray}\label{pdf grid}
\bar{\rho}(x,y;s)=\bar{f}(x,s)\exp\left(-\frac{|y|}{\sqrt{\mathcal{D}_{y}\bar{\mathcal{K}}_{y}(s)}}\right).
\end{eqnarray}
Note that the exponential term carries information about the diffusion in the branches. From Eq.~(\ref{pdf grid}), we find a relation for the probability distribution of the $x$-direction in the Laplace domain, that is,
\begin{eqnarray}
\label{Green grid}
\bar{\mathcal{G}}_{1}(x;s)=2\sqrt{\mathcal{D}_{y}\bar{\mathcal{K}}_{y}(s)}\bar{f}(x,s).
\end{eqnarray}
On the other hand, we also have that $\bar{\mathcal{G}}_{1}(x,t)=\int_{-\infty}^{\infty}dy\,\bar{\rho}(x,y,t)$ and by integrating Eq.~(\ref{comb laplace2}), we find
\begin{eqnarray}\label{comb laplace2 backbone}
s\bar{\mathcal{G}}_{1}(x;s)-\Phi(x) = s{\cal{D}}_{x}\bar{\mathcal{K}}_{x}(s)\frac{\partial^2}{\partial x^2} \sum_{l_{j}\in\mathcal{S}_{\nu}}\bar{\rho}(x,y=l_{j};s) \;,
\end{eqnarray}
where $\Phi(x)=\int_{-\infty}^{\infty}dy\,\varphi(x,y)$, and $\bar{\rho}(x,y=l_{j};s)$ represents the probability distribution function for the backbone in the position $l_{j}$. 

The summation $\sum_{l_{j}\in\mathcal{S}_{\nu}}\bar{\rho}(x,y=l_{j};s)$ can be formally replaced by integration to fractal measure $\mu_{\nu}\sim l^{\nu}$, where $\sum_{l_j\in\mathcal{S}_{\nu}}\delta(l-l_{j})\rightarrow\frac{1}{\Gamma(\nu)}l^{\nu-1}$ is the fractal density, that is, $d\mu_{\nu}=\frac{1}{\Gamma(\nu)}l^{\nu-1}\,dl$ \cite{tarasov}. Therefore, from Eqs.~(\ref{pdf grid}) and (\ref{Green grid}), we obtain  
\begin{eqnarray}\label{fractal summation}
\sum_{l_{j}\in\mathcal{S}_{\nu}}\bar{\rho}(x,y=l_{j};s)&=&\sum_{l_{j}\in\mathcal{S}_{\nu}}\bar{f}(x,s)\exp\left(-\frac{|l_{j}|}{\sqrt{\mathcal{D}_{y}\bar{\mathcal{K}}_{y}(s)}}\right)\nonumber\\
&=&\bar{f}(x,s)\frac{1}{\Gamma(\nu)}\int_{0}^{\infty}dl\,l^{\nu-1}e^{-\frac{l}{\sqrt{\mathcal{D}_{y}\bar{\mathcal{K}}_{y}(s)}}}\nonumber\\
&=&\bar{f}(x,s)\left[\mathcal{D}_{y}\bar{\mathcal{K}}_{y}(s)\right]^{\nu/2}=\frac{1}{2\mathcal{D}_{y}^{\frac{1-\nu}{2}}}\left[\bar{\mathcal{K}}_{y}(s)\right]^{-\frac{1-\nu}{2}}\bar{\mathcal{G}}_{1}(x;s).
\end{eqnarray}
We observe in Eq.~(\ref{fractal summation}) that the power-law exponents of the diffusion coefficient and memory kernel of $y$-direction contains all information about the fractal structure of backbones. Hence, by using the result of Eq.~(\ref{fractal summation}) in Eq.~(\ref{comb laplace2 backbone}), we find
\begin{eqnarray}\label{comb laplace2 backbone2}
s\bar{\mathcal{G}}_{1}(x;s)-\Phi(x) = \frac{\mathcal{D}_{x}}{2\mathcal{D}_{y}^{\frac{1-\nu}{2}}}s\frac{\bar{\mathcal{K}}_{x}(s)}{\left[\bar{\mathcal{K}}_{y}(s)\right]^{\frac{1-\nu}{2}}}\frac{\partial^2}{\partial x^2}\bar{\mathcal{G}}_{1}(x;s).
\end{eqnarray}

The inverse Laplace transform of Eq.~(\ref{comb laplace2 backbone2}) yields the generalized diffusion equation
\begin{eqnarray}\label{comb2 backbone2}
\frac{\partial}{\partial t}\mathcal{G}_{1}(x;t) = \left(\frac{\mathcal{D}_{x}}{2\mathcal{D}_{y}^{\frac{1-\nu}{2}}}\right)\frac{\partial}{\partial t}\int_{0}^{t}dt'\,\eta(t-t')\frac{\partial^2}{\partial x^2}\mathcal{G}_{1}(x;t'),
\end{eqnarray}
where the memory kernel $\eta(t)$ depends on both annealed and quenched disorder mechanisms, as given by
\begin{eqnarray}
\eta(t)=\mathcal{L}^{-1}\left\{\frac{\bar{\mathcal{K}}_{x}(s)}{\left[\bar{\mathcal{K}}_{y}(s)\right]^{\frac{1-\nu}{2}}}\right\}.
\end{eqnarray}
By using Eq.~(\ref{comb2 backbone2}), we obtain the general expression for the MSD of a comb structure related to the fractal dimension $0<\nu<1$ in the Laplace domain, that is, 
\begin{eqnarray}\label{msd general fractal mesh}
\sigma_{x}^{2}(t)=2\left(\frac{\mathcal{D}_{x}}{2\mathcal{D}_{y}^{\frac{1-\nu}{2}}}\right)\mathcal{L}^{-1}\left\{\frac{\bar{\mathcal{K}}_{x}(s)}{s[\bar{\mathcal{K}}_{y}(s)]^{\frac{1-\nu}{2}}}\right\}.
\end{eqnarray}
We note again that probability distributions along the backbones structure should be non-negative; therefore: $1/\left[s\bar{\eta}(s)\right]=\left[\bar{\mathcal{K}}_{y}(s)\right]^{\frac{1-\nu}{2}}/\left[s\bar{\mathcal{K}}_{x}(s)\right]$ should be completely monotone function, and $1/\bar{\eta}(s)=\left[\bar{\mathcal{K}}_{y}(s)\right]^{\frac{1-\nu}{2}}/\bar{\mathcal{K}}_{x}(s)$ should be a Bernstein function.

To investigate the isolated effects of geometrical restrictions of the fractal grid, we consider the case $\bar{\mathcal{K}}_{x}(t)=\bar{\mathcal{K}}_{y}(t)=1$ (that is,  $\bar{\mathcal{K}}_{x}(s)=\bar{\mathcal{K}}_{y}(s)=1/s$) where the MSD recovers the result of Ref.~\cite{pre2015}, that is,
\begin{eqnarray}
\label{MSD2015}
\sigma_{x}^{2}(t)=2\left(\frac{\mathcal{D}_{x}}{2\mathcal{D}_{y}^{\frac{1-\nu}{2}}}\right)\mathcal{L}^{-1}\left\{\frac{s^{-1}}{s^{1-\frac{1-\nu}{2}}}\right\}=2\frac{\mathcal{D}_{x}}{2\mathcal{D}_{y}^{\frac{1-\nu}{2}}}\frac{t^{\frac{1+\nu}{2}}}{\Gamma\left(1+\frac{1+\nu}{2}\right)}.
\end{eqnarray}
The power-law behavior ($\sigma_{x}^{2}(t)\simeq t^{\frac{1+\nu}{2}}$) indicates a scale invariance, and because the exponent depends on $\nu$, we infer that the fractal grid affects the spreading dynamics in all time scales. The overall effect is a subdiffusive dynamics with $\frac{1}{2}<\frac{1+\nu}{2}<1$; however, the subdiffusion in a fractal grid is faster than the usual comb-model ($\sigma_{x}^{2}(t)\simeq t^{1/2}$~\cite{ArBa}, corresponding to $\nu=0$ in our case). This behavior occurs because the set of backbones increases the possibilities of diffusion in the $x$-direction. For example, a fractal grid with $\nu \simeq 0.631$ (the one-third Cantor set) implies in $\sigma_{x}^{2}(t)\simeq t^{0.816}$. Moreover, Eq.~(\ref{MSD2015}) connects the anomalous diffusion exponent with the fractal dimension of the backbone structure, a result that was experimentally observed in diffusion in porous and structurally inhomogeneous media~\cite{zhokh}. 

One example of the interplay between this generalized geometrical restriction and the effect of memory kernels is given by $\bar{\mathcal{K}}_{x}(s)=\bar{\mathcal{K}}_{y}(s)=1/s^{\alpha}$. This choice yields a fractional diffusion equation for a fractal grid given by
\begin{eqnarray}
\label{Comb_fract alpha}
\frac{\partial}{\partial t}\rho(x,y;t) = {_{RL}}D_{t}^{1-\alpha}\left[
{\cal{D}}_{y}\frac{\partial^2}{\partial y^2} \rho(x,y;t) +
{\cal{D}}_{x}\sum_{l_{j}\in\mathcal{S}_{\nu}}\delta(y-l_{j})\frac{\partial^2}{\partial x^2} \rho(x,y;t)\right]\,,
\end{eqnarray}
and whose MSD is
\begin{eqnarray}
\sigma_{x}^{2}(t)=2\left(\frac{\mathcal{D}_{x}}{2\mathcal{D}_{y}^{\frac{1-\nu}{2}}}\right)\mathcal{L}^{-1}\left\{s^{-\frac{\alpha+\nu}{2}-1}\right\}=2\left(\frac{\mathcal{D}_{x}}{2\mathcal{D}_{y}^{\frac{1-\nu}{2}}}\right)\frac{t^{\frac{\alpha+\nu}{2}}}{\Gamma\left(\frac{\alpha+\nu}{2}\right)}.
\end{eqnarray}
The power-law exponent $\frac{\alpha+\nu}{2}$ is associated with the two anomalous diffusion mechanisms: memory effects (related to $\alpha$) and the fractal structure restriction (given by $\nu$). The interplay between these mechanisms produces subdiffusive regimes between the limit cases of restricted and Brownian diffusion, that is, $0<\frac{\alpha+\nu}{2}<1$. The case of a single backbone ($\nu=0$) yields $\sigma_{x}^{2}(t)\simeq t^{\alpha/2}$, as reported in Ref.~\cite{mmnp}. From the general formula of Eq.~(\ref{msd general fractal mesh}), we further conclude that the MSD along the $x$-direction is stationary (case of localization) if  $\bar{\mathcal{K}}_{x}(s)\propto\left[\bar{\mathcal{K}}_{y}(s)\right]^{\frac{1-\nu}{2}}$, and that normal diffusion along the $x$-direction occurs for $\bar{\mathcal{K}}_{x}(s)\propto s^{-1}\left[\bar{\mathcal{K}}_{y}(s)\right]^{\frac{1-\nu}{2}}$. A remarkable feature of these conditions is that both memory kernels must have the same effects of geometrical restrictions.

\section{Discussion and Conclusions}~\label{sec:4}
Over this work, we showed that our generalized comb-models account for annealed and quenched disorder mechanisms. We believe these comb-models with fractional time-derivative operators are a reasonable abstraction for systems where the interplay between temporal and spatial disorders is present. For these hybrid models, we considered different memory kernels for the backbone and branch structures, and a fractal generalization of the geometrical restrictions. We obtained general solutions for the diffusion propagator and the MSD in terms of these memory kernels. With these solutions, we discussed particular cases based on the temporal evolution of the MSD and inferred time scales associated with each disorder mechanism. These results thus provide theoretical knowledge about the importance of interactions between geometrical restrictions and memory effects on anomalous diffusion.  

We argued that the behaviors obtained from our models are consistent with other theoretical and experimental results. In its usual form, the comb-model is a subdiffusive model with $\sigma_{x}^{2}\propto t^{1/2}$. However, depending on the memory kernels and number of backbones (single or fractal set), our generalized comb-models also describe restricted diffusion, Brownian diffusion, and display a crossover from subdiffusion to these situations. 

For power-law memory kernels, the MSD is scale-invariant and the diffusive regime depends on the values of memory exponents $0<\alpha_{x}<1$ and $0<\alpha_{y}<1$ with $\alpha_{x}>\alpha_{y}/2$. Scale invariance is also a feature of the fractal grid structure of backbones. In this case, there is an enhancement of the diffusion (when compared with the usual comb-model) and the anomalous exponent depends on the fractal dimension ($\frac{1}{2}<\frac{1+\nu}{2}<1$). By including power-law memory effects on the fractal grid, we obtained an anomalous exponent $0<\frac{\alpha+\nu}{2}<1$ with $\alpha= \alpha_{x}=\alpha_{y}$. Overall, these results from our generalized comb-models are consistent with simulations of fractal Brownian motion~\cite{Ribeiro2014} and a biased CTRW~\cite{Mendez2015} on comb-like structures. 

When the memory kernel acting on the branches is exponential, we showed that the behavior of the MSD in the $y$-direction is similar to those reported in diffusion with explicit confinement (external forces or limited boundaries) or having a bias to return to the origin. The spread in the branches thus follows Brownian diffusion for short-time scales and becomes confined near to the backbone for long times. Because of this high probability of returning to the backbone, the long-time behavior in $x$-direction only depends on its memory kernel. The overall effect of an exponential kernel is a crossover between diffusive regimes. This crossover occurs from subdiffusion to Brownian diffusion when walkers have no memory in the backbone. Thus, we observed that the interplay between two subdiffusion mechanisms (geometrical restrictions and memory effects) may lead to the usual Brownian motion. This crossover and its physical explanation also appear consistent with experimental results of anisotropic diffusion of entangled biofilaments~\cite{Tsang2017} and the mobility of water molecules on protein surfaces~\cite{HongPRL2018}. On the other hand, when an exponential memory kernel acts on the backbone, we obtained a crossover from subdiffusion to confined diffusion. This result also appears consistent with the MSD observed in lateral diffusion of membrane receptors~\cite{kusumi1993} and diffusion of protein aggregates in live {\it E. coli} cells \cite{Coquel2013}. Moreover, we found that the confinement effects of exponential memory kernels are independent of the parameters $\alpha'_{y}$ and $\alpha'_{x}$. 

In spite of its simplicity, our generalized comb-model may have an important role in the statistical mechanics of disordered media. This model can be used as a simple explanation for unusual transport properties caused by quenched disorder or to annealed disorder. It also has the advantage of providing exact solutions related to sub- and superdiffusive behaviors. Our comb-models are also relevant for investigating the interplay between temporal and spatial disorder mechanisms and for describing crossovers from subdiffusion to confined or Brownian diffusion.

\section*{Acknowledgments}

E.K.L. thanks partial financial support of the CNPq under Grant No. 302983/2018-0. EKL also thanks the National Institutes of Science and Technology of Complex Systems -- INCT-SC for partially supporting this work. HVR thanks the financial support of the CNPq under Grants 407690/2018-2 and 303121/2018-1. TS and IP acknowledge the support by the bilateral research project MK 07/2018, WTZ Mazedonien S$\&$T Macedonia 2018--20 funded under the inter-governmental Macedonian-Austrian agreement. TS was supported by the Alexander von Humboldt Foundation.

\linespread{1.01}
\bibliography{fractional-comb}

\end{document}